\documentclass{elsart}
\usepackage{graphicx}
\usepackage{amssymb}
\begin{document}

\begin{frontmatter}

\title{
Destruction of the family of steady states in the planar problem of  Darcy convection
}

\author{Tsybulin V.~G.\thanksref{label3}
,}
\author{
Karas\"{o}zen B.
\thanksref{label2}
}
\ead{bulent@metu.edu.tr}
\corauth[cor1]{Ankara, Turkey}
\address[label3]{Department of Computational Mathematics,
\\
Southern Federal University, Rostov on Don,
 Russia}
\address[label2]{Department of Mathematics and Institute of Applied Mathematics,
\\
Middle East Technical University, Ankara, Turkey}

%
%
%

\begin{abstract}
The natural convection of incompressible fluid in a porous medium  causes for some boundary conditions a strong non-uniqueness in the form of a continuous family of steady states.
We are interested in the situation when these boundary conditions are violated.
The resulting destruction of the family of steady states is studied via computer experiments based on a mimetic finite-difference approach.
Convection in a rectangular enclosure is considered under different perturbations of boundary conditions (heat sources, infiltration).
Two scenario of the family of equilibria are found: the transformation to a limit cycle and the formation of isolated convective patterns.
\end{abstract}

\end{frontmatter}


\setcounter{equation}{0}

\section{Introduction}
\label{sec1}

The Darcy model of two-dimensional incompressible fluid convection
in a porous medium \cite{Lyu75,GluLyPut78,Yud91,Yud95} demonstrates
the nontrivial effect of the appearance of a continuous family of
steady states (equilibria) with a variable spectrum of stability. It
was explained in \cite{Yud91} that the cosymmetry can lead to  this
phenomenon. Cosymmetry differs from symmetry due to variability of
spectrum where all members of the family have the identical spectrum
of stability.
 The effects of cosymmetry in Darcy  convection was analyzed theoretically in \cite{Yud91,Yud95,Yud04}.
The computation of the families of steady states and their evolution
was carried out in \cite{Gov99,KarTsy99,Gov00,KanTsy03,KarTsy04,KarTsy05b}.
Partial instability of the family (the occurrence of unstable steady states) was studied in \cite{Gov99,Gov00,KanTsy03}.
The effects of collision and reorganization of the families were analyzed in \cite{Gov00,KanTsy03,KarTsy04}.

A related  interesting phenomenon  is the destruction of a family of equilibria  under some perturbations \cite{BraLyuRou95,Yud04,KurYud04}.
In \cite{BraLyuRou95} weak  seeping of the fluid through the boundaries was considered.
The breakdown of the family of stationary solutions was analyzed using the normal forms technique.
V. Yudovich  \cite{Yud04}  studied the violation of the cosymmetry under low-intense heat sources or weak fluid infiltration and found that the continuous family of equilibria  disintegrates or even disappears.
It was predicted that in the case of heat sources  a family of steady states may disintegrate to produce several separate stationary regimes.
The case of weak infiltration may lead to the appearance of a long-period limit cycle instead of the family of equilibria \cite{BraLyuRou95,Yud04}.

In this paper we study   the problem of the destruction of the family of steady states in  the planar problem of Darcy convection in a  rectangle.
The finite-difference approximation of the Darcy equation allows us to consider the perturbation listed in \cite{Yud04} but our numerical studies are  not restricted to the cases of small disturbances of heat source and low velocity of infiltration.
We carried out several computer experiments with the harmonic perturbations of the temperature on the bottom (the case of a finite number of heat sources located on the boundary).
The disintegration of the family of continuous equilibria is demonstrated numerically by showing the appearance of limit cycles  and the formation of  isolated convective patterns.

The paper is organized as follows.
In Section 2 we give the Darcy equation and the cosymmetry property briefly.
In Section 3 the finite-difference discretization is outlined.
In Section 4 we present the numerical results for different perturbations of the boundary conditions.

\section{Darcy model and cosymmetry}
We consider an incompressible fluid which saturates a  porous medium in a rectangular enclosure $ \mathcal{D} = [0,a] \times [0,b]$ which is heated below.
The dimensionless equations of the fluid flow in a porous medium according to Darcy law can be written as \cite{Yud95}:
\begin{eqnarray}
\label{eq_theta}
   &&  \theta_t = \Delta \theta + \lambda \psi_x + J (\theta,\psi)
\equiv F
\\
\label{eq_psi}
   &&  0  = \Delta \psi - \theta_x \equiv G ,
\end{eqnarray}
where $\Delta $ is the Laplacian and $J(\theta,\psi)$ denotes the Jacobian  operator over $(x,y)$
$
J(\theta,\psi) = \theta_x \psi_y -\theta_y \psi_x
$.
The unknown functions $\psi(x,y,t)$ and $\theta(x,y,t)$ denote, respectively,  the stream function and the perturbation of temperature from the basic state of rest with a linear conductive profile, $\lambda$ is the Rayleigh number.

The boundary conditions are:
\begin{eqnarray}
\label{bound}
   &&    \theta(x,0,t)=\eta f(x); \quad \theta(x,y,t)=0, \quad (x,y) \in \partial \mathcal{D} \setminus \{y=0\} ;
\\
\nonumber
&& \psi(x,y,t) = \mu y, \quad (x,y) \in \partial \mathcal{D},
\end{eqnarray}
here the function $f(x)$ gives the nonhomogeneous distribution of the temperature at the bottom with  magnitude  $\eta$,  and $\mu$ is the velocity of infiltration through the lateral sides of an enclosure.

The problem is supplied by the initial conditions
\begin{equation}
\label{init}
  \theta (x,y,0) = \theta_0 (x,y),
\end{equation}
where $\theta_0$ denotes the initial perturbation of the temperature from the linear profile.
 For a given $\theta_0$, the stream function $\psi$ can be obtained from (\ref{eq_theta})--(\ref{bound}) as the solution of the Dirichlet problem via Green's operator.

When  $\eta=\mu=0$ the system (\ref{eq_theta})--(\ref{bound})  imposes the equilibrium  $\theta=\psi=0$.
This state of rest remains stable while $\lambda < \lambda_{11}$.
The critical values $\lambda_{ij}$ are given by
$\lambda_{ij} = \pi^2 ( i^2/a^2 + j^2/b^2)$,
$ i, j \in Z$.
It was shown in \cite{Yud95}  that the first critical value $\lambda_{11}$ is two-times for any domain $\mathcal{D}$ and larger values of $\lambda$ implies an onset of the family of steady states.
This result is a consequence of the cosymmetry that exists for the system (\ref{eq_theta})--(\ref{bound}) when $\eta=\mu=0$ \cite{Yud91}.
Indeed, the vector-function $(\theta, -\psi)$ is orthogonal to the right-hand side of (\ref{eq_theta}), (\ref{eq_psi}) in $L_2$, i.e $ \int_{ \mathcal D} (F \psi - G \theta) dx dy =0$.
In the case of nonzero $\eta$ and $\mu$ we obtain using integration by parts and Green's formulae
\begin{eqnarray*}
 \int_{ \mathcal D} (F \psi - G \theta) dx dy
  =   & - &\eta \int_0^a \psi_y(x,0,t)f(x) dx
- \mu b \int_0^a \theta_y(x,b,t) dx
 \\
& - & \mu \int_0^b y [\theta_x(a,y,t) - \theta_x(0,y,t)] dy
.
\end{eqnarray*}
 Thus, the cosymmetry condition fails whenever $\mu \ne 0$ or $\eta \ne 0$.

\section{Finite difference discretization}
We use here the discretization and numerical scheme derived in \cite{KarTsy05a}.
Let us introduce the uniform arrangement of the nodes on the interval for spatial coordinates:
\begin{eqnarray*}
&& x_n=nh_x, \quad n=0 ,\cdots, N+1, \quad h_x=a/(N+1),
\\
&& y_m=mh_y, \quad m=0 ,\cdots, M+1, \quad h_y=b/(M+1) .
\end{eqnarray*}
The temperature  $\theta$ and stream function $\psi$ are defined at the nodes $(x_n,y_m)$.
It is useful to introduce the auxiliary (staggered) nodes on each of the coordinates
$x_{n+1/2} = \frac12 (x_n + x_{n+1})$, $n=0 ,\cdots, N$, and $y_{m+1/2} = \frac12 (y_m + y_{m+1})$, $m=0 ,\cdots, M$.

To approximate equations (\ref{eq_theta})--(\ref{eq_psi}) we introduce difference and  averaging operators based on the two-node stencils
\begin{eqnarray}
\label{delta_x}
  (\delta_x f)_{n+1/2,m}  =
  \frac{f_{n+1,m} - f_{n,m}}{ h_x}
,
\quad
  (\delta_y f)_{n,m+1/2}  =
  \frac{f_{n,m+1} - f_{n,m}}{ h_y}
,
\\
\label{delta_0}
(\delta_{0x} f)_{n+1/2,m}  =
\frac{  f_{n+1,m}
      + f_{n,m} }
     { 2 } ,
\quad
(\delta_{0y} f)_{n,m+1/2} =
\frac{ f_{n,m+1}
      + f_{n,m}  }
     { 2} .
\end{eqnarray}
These formulas are valid for integer and half-integer values of $n$ and $m$.
Using operators (\ref{delta_x})--(\ref{delta_0}) we derive  the operators on three-node stencils
\begin{equation}
\label{D1D2}
 D_x   =   \delta_x \delta_{0x} \equiv  \delta_{0x} \delta_{x} ,
  \quad
 D_y f  =  \delta_y \delta_{0y} \equiv  \delta_{0y} \delta_{y}
,
\end{equation}
the averaging and difference operators on a four-point stencil
\begin{equation}
\label{d0_dx_dy}
d_0   =  \delta_{0x} \delta_{0y} \equiv  \delta_{0y} \delta_{0x} ,
\quad
d_x  = \delta_{x} \delta_{0y} \equiv  \delta_{0y} \delta_{x} ,
\quad
d_y = \delta_{y} \delta_{0x} \equiv  \delta_{0x} \delta_{y} ,
\end{equation}
and the discrete analog of Laplacian
\begin{equation}
\label{Delta}
(\Delta_h f)_{n,m} =
\left( \delta_x \delta_x f + \delta_y \delta_y f \right)_{n,m}.
\end{equation}

To approximate  the Jacobian we use the Arakawa scheme \cite{Ara66}
\begin{eqnarray}
\label{jac}
J_{n,m}(\theta,\psi)
= && \frac13 \big[
   D_x \left( \theta D_y \psi \right)
 - D_y \left( \theta D_x \psi \right)
\big]_{n,m}
\\
\nonumber
 &+& \frac23
\big[ d_x \left( d_0 \theta d_y \psi \right)
     - d_y \left( d_0 \theta d_x \psi \right)
\big]_{n,m} .
\end{eqnarray}

The discretized boundary conditions are formulated as follows
\begin{eqnarray}
\label{bc_fd}
&&
\psi_{n,0} = 0, \quad
\theta_{n,0}  =  \eta f(x_n), \quad
n =\ 0,\ldots, N+1,
\\
\nonumber
&&
\psi_{n,M+1} = 0, \quad
\theta_{n,M+1} = 0 , \quad
n =\ 0,\ldots, N+1,
\\
\nonumber
&&
\theta_{0,m}  =  \theta_{N+1,m} =0,
\enskip \psi_{0,m} = \psi_{N+1,m} = \mu y_m, \quad
m =\ 0,\ldots, M+1 .
\end{eqnarray}

Using operators (\ref{D1D2})--(\ref{jac})  we obtain the  semi-discretized form of (\ref{eq_theta}) and (\ref{eq_psi}):
\begin{eqnarray}
\label{ode1}
&&
\dot{\theta}_{nm}  =  F_{nm} \equiv
  \left( \Delta _h \theta + \lambda D_x \psi \right)_{nm}
   + J_{nm}(\theta,\psi) ,
\\
\label{ode2}
&&
0  =  G_{nm} \equiv
  \left( \Delta _h \psi - D_x \theta \right)_{nm},
\end{eqnarray}
here $n=1,\cdots, N$, $m=1 ,\cdots, M$, the dot denotes a differentiation on $t$.

%

After discretization the resulting system of ordinary differential equations (\ref{ode1})--(\ref{ode2}) may be written in the following form
\begin{equation}
\label{ode}
\dot \Theta = ( A+\lambda B A^{-1} B ) \Theta  +F(\Theta)
 = \Phi(\Theta,\lambda,\mu,\eta).
\end{equation}
Here $\Theta$ is vector of unknowns (temperature), matrices $A$ and $B$ are  the approximation, respectively, of the  Laplacian $\Delta$ and the first derivative on $x$, where $F(\Theta)$ represents the nonlinear Jacobian term.

To find nonstationary regimes or to converge to stable equilibrium we apply the Runge-Kutta method of fourth order.
To compute a family of equilibria in the system (\ref{ode}), we apply the technique \cite{KarTsy99} based on the cosymmetric version of the implicit function theorem \cite{Yud95}.
Different realization for ordinary differential equations  \textit{was also} developed in \cite{Gov99}.

When $\lambda$ is slightly larger than $\lambda_{11}$,  then all points of the family are stable \cite{Yud95}.
Starting from the vicinity of unstable zero equilibrium we integrate the ordinary differential equations (\ref{ode}) by \textit{the} Runge-Kutta method up to a point $\Theta_0$ close to some stable equilibrium on the family.
Then we correct the point $\Theta_0$ by the Newton method.
To predict the next point on the family we determine the kernel of the linearization matrix (Jacobi matrix) at the point $\Theta_0$ and use the Adams-Bashford method.
We repeat this procedure to obtain the entire family.

\section{Numerical results}
Here we mainly consider  convection for a wide enclosure with $a=2$ and $b=1$.
Our previous computation of the families of steady states \cite{KarTsy99,KarTsy04} showed that a sufficiently good description of a number of the transitions can be obtained on a rather coarse grid.
Several hundreds of internal nodes was enough to describe an appearance of  a primary family of steady states \cite{KarTsy99} and follow a number transformations \cite{KarTsy04}.
So we used the discretization with  $16 \times 8$ nodes and compared with results on $32 \times 16$ nodes
 to study the destruction of a family of steady states under inhomogeneity of boundary conditions.

To present results we apply the special Nusselt coordinates \cite{Gov99}
\begin{equation}
\label{Nu}
Nu_h  =  \int_{0}^b \theta_x(\frac a2,y) dy,
\quad
Nu_v  =  \int_{0}^a \theta_y(x,0) dx ,
\end{equation}
where $Nu_h$ corresponds to the integral value of the heat flux from left to  right defined for the centered vertical section.
The value $Nu_v$ may be considered as a cumulative heat flux through the bottom of the container.

\subsection{Infiltration through the lateral sides}
In  \cite{BraLyuRou95,Yud04}  a scenario of the transformation of a family of steady states to a limit cycle of the large period was described.
\begin{figure}[htb]
\centerline{
\includegraphics[scale=0.7]{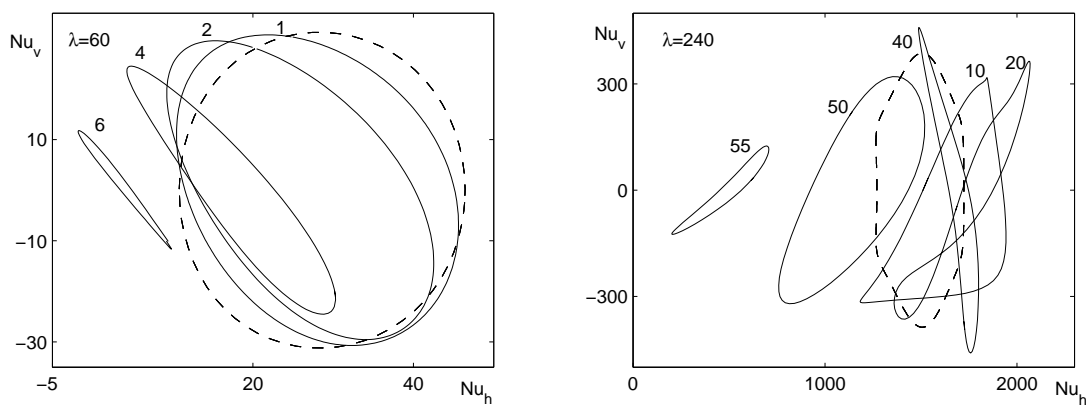}
}
\caption{
Family of steady states (dashed) and limit cycles (solid) for several infiltration velocities  $\mu$:
$\lambda=60$ (left) è $\lambda=240$ (right)
}
\label{fam_cycles_168_2_1_60_240}
\end{figure}
Fig.~\ref{fam_cycles_168_2_1_60_240} illustrates  this scenario for two Rayleigh numbers
in the case of small and large infiltration through the lateral sides.
The case with  $\lambda=60$ corresponds to the initial stage of the evolution of the family of equilibria after its birth.
The  Rayleigh number $\lambda=240$ shows the rather developed convective patterns.
If infiltration velocity $\mu$ is small then the trajectory of a limit cycle lies in vicinity of the family of equilibria.
When the velocity $\mu$ increases a limit cycle goes far away from the family.
Moreover, large infiltration causes degradation of the nonstationary convective regime.
All fluid particles move horizontally without the formation of a convective pattern.
With the increase of the Rayleigh parameter $\lambda$ the convective regime becomes able to  prevent a pure infiltration.
Computation at $\lambda=240$ shows that for small velocity  $\mu$ a limit cycle occurs near a curve of the family of equilibria.
Increasing of $\mu$ shifts \textit{the limit cycle to the range of small Nusselt values} $Nu_h$.
Afterwards we see a deformation of a limit cycle and rather large velocity of infiltration leads to suppression of the convection.

\begin{figure}[htb]
\centerline{
\includegraphics[scale=0.7]{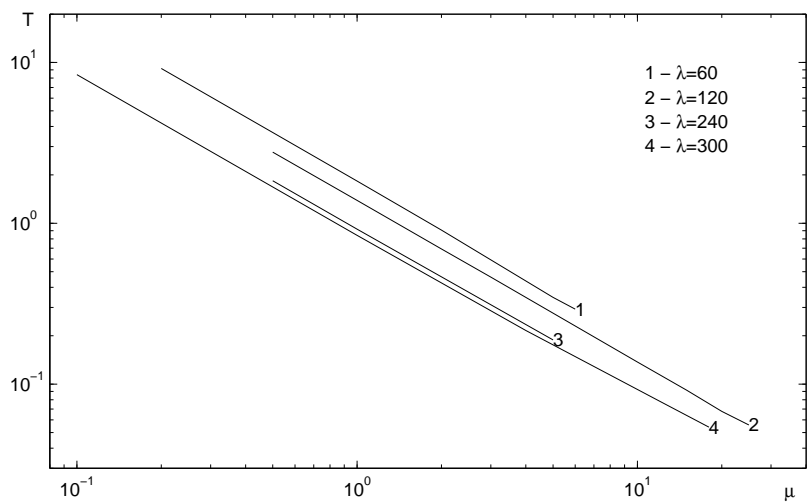}
}
\caption{
Dependence of the period of a limit cycle from the velocity of infiltration $\mu$ for several $\lambda$
}
\label{T_vs_mu_168_2_1_60_120_240_300}
\end{figure}

The period of a limit cycle  $T$  becomes smaller with increasing  velocity $\mu$ and depends also on Rayleigh parameter $\lambda$.
One can see in Fig.~\ref{T_vs_mu_168_2_1_60_120_240_300} that the dependence $T(\mu)$ is practically the same for different values of $\lambda$.

\subsection{Inhomogeneous heating from below}
For  inhomogeneous heating and without infiltration ($\mu=0$) we found two scenarios of destruction of the family of equilibria.
\begin{figure}[htb]
\centerline{
\includegraphics[scale=0.75]{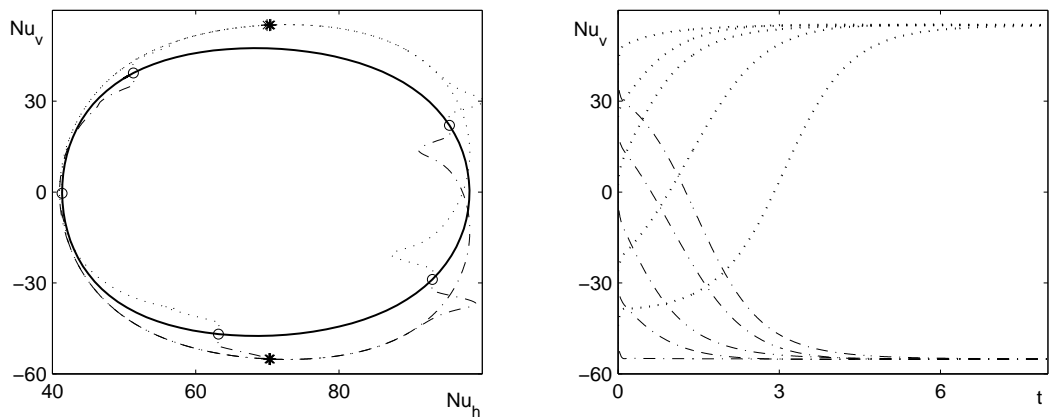}
}
\caption{
Family of steady states and convergence from different initial points (circles) to the isolated stationary states (stars) under inhomogeneous heating below: $\eta=4$ (dashed), $\eta=-4$ (dashed-point);
$\lambda=70$, $\kappa=4$
}
\label{Isol_equi_a2b1_70_Eta0_4_kap_4}
\end{figure}
We consider the case of the harmonic distribution of the bottom temperature $f(x)= sin(\kappa \pi x/a)$.
It allows us to consider the case with zero average deviation from  uniform heating.
Through the experiment we change the intensity of heating (Rayleigh number $\lambda$), the magnitude of inhomogeneity  $\eta$ and the amount of half-harmonics $\kappa$.
Depending on the  parameters, the family of steady states may transform to a limit cycle or to a number of isolated equilibria.

Computation with two harmonics in temperature distribution ($\kappa=4$) shows that the family of steady states disintegrates onto two isolated equilibria, see Fig.~\ref{Isol_equi_a2b1_70_Eta0_4_kap_4}.
One of these equilibria is stable while the second is unstable,
it depends on the sign of magnitude  $\eta$.

The computer experiment was started from  points  taken from the curve of the family of equilibria.
For small values of magnitude  $\eta$ the trajectories were very close to the curve of the family of equilibria.
In this case we found very slow motion to the isolated stationary state.

\begin{figure}[htb]
\centerline{
\includegraphics[scale=0.75]{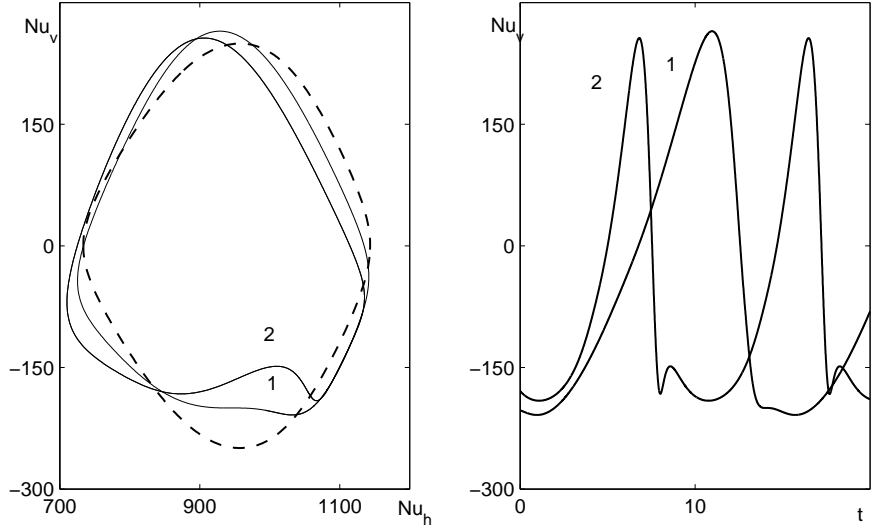}
}
\caption{
The limit cycle (dashed) and the family of steady states (solid); $\lambda=190$,  $\eta=8$ (curve 1), $\eta=16$ (curve 2), $\kappa=2$
}
\label{Dest_LC_a2b1_190_Eta0_8_kap_2}
\end{figure}

In the case of one harmonics ($\kappa=2$) the scenario of destruction of a family onto isolated equilibria was observed when $\lambda<180$. The transformation of the family of steady states to a limit cycle  was found for $\lambda \ge 180$.
Fig.~\ref{Dest_LC_a2b1_190_Eta0_8_kap_2} gives relative location of the family of steady states and two limit cycles.
For small magnitudes $\eta$  the trajectory of the limit cycle lies close to the family.

\begin{figure}[htb]
\centerline{
\includegraphics[scale=0.75]{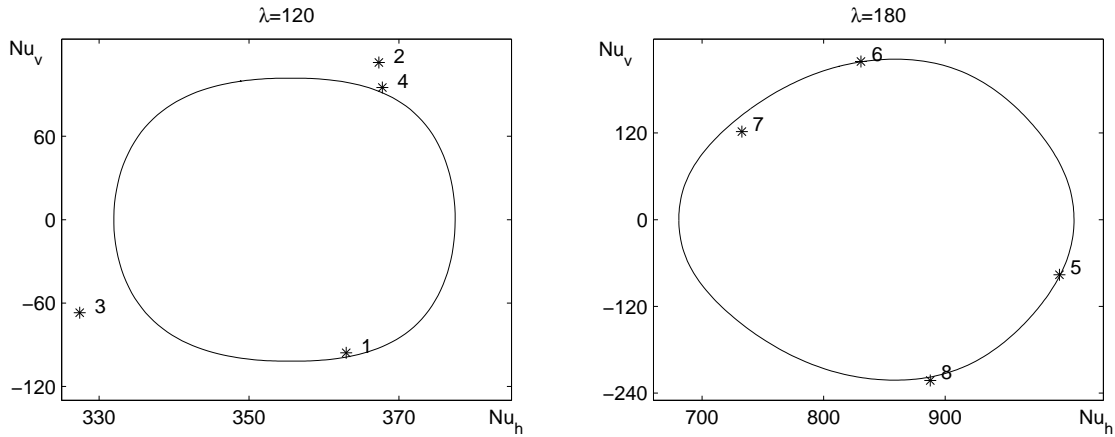}
}
\caption{
Families of steady states (solid) and isolated states (stars) for heating with magnitude of perturbation $\eta=8$ and different number of half-harmonics $\kappa$.
Left: $\lambda=120$, $\kappa=2$ (1), $\kappa=4$ (2), $\kappa=6$ (3), $\kappa=6$ (4);
Right: $\lambda=180$; $\kappa=2$ (5),  $\kappa=4$ (6), $\kappa=6$ (7), $\kappa=6$ (8)
}
\label{family_168_120_2_1_E8K246}
\end{figure}

\begin{figure}[htb]
\centerline{
\includegraphics[scale=0.75]{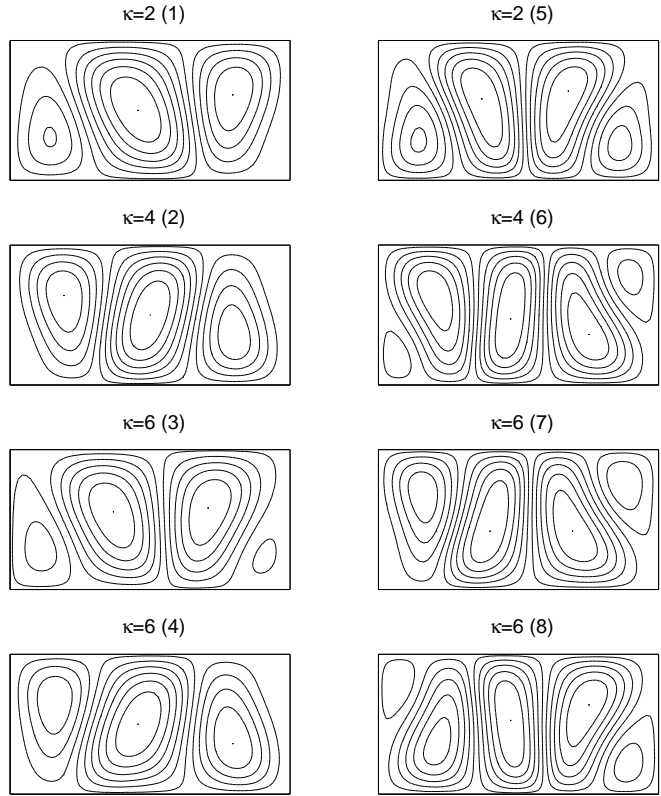}
}
\caption{
Stream function of isolated steady states for different $\kappa$: $\lambda=120$ (left), $\lambda=180$ (right); $\eta=8$
}
\label{states_168_120_2_1_E8K246}
\end{figure}

The convective patterns and their number  depend on the number of half-harmonics $\kappa$.
Fig.~\ref{family_168_120_2_1_E8K246} shows the results of disintegration of the families for two Rayleigh numbers:  $\lambda=120$ and $\lambda=180$.
We changed $\kappa$ and fixed  a magnitude of heating perturbation $\eta=8$.
For $\kappa=2$ and $\kappa=4$ only one isolated equilibrium was observed.
The case $\kappa=6$ produces two isolated stable equilibria instead of the family of steady states.
The unstable equilibria also exist and may be obtained using computation with $\eta=-8$.
Here we see a change of stability properties for isolated equilibria to which the family of steady states transforms.
Stream functions for realized convective patterns are given in  Fig.~\ref{states_168_120_2_1_E8K246}.
It is clearly seen that one state for $\kappa=6$ looks as if the equilibrium being formed under temperature distribution with two harmonics ($\kappa=4$).
The second possible state for $\kappa=6$ is rather different.
These experiments show that we can control the desirable state by the appropriate choosing of magnitude and amount of harmonics.

\section{Conclusion}

We have studied convection in the rectangular enclosure under the
following perturbations of boundary conditions: heat sources at the
bottom and infiltration through the lateral sides. Nonzero boundary
values destroy the family of steady states which existed in the case
of Dirichlet boundary conditions. Computer experiment allows to
consider the case of large magnitudes of perturbations. Two scenario
were studied: transformation of the family to a limit cycle and
destruction of a family on a number of isolated convective patterns.

We found through computation that convergence to the isolated steady states may be very slow.
It looks as a memory effect when the system remembers the location of the former family of equilibria.
It should be noted that the dynamics on finite time span is very close to the results on selection of equilibria under bottom heat sources \cite{TsyKarErg06}.
This may be used to control the preferable convective state via variation of boundary conditions.

\subsection*{Acknowledgements}
The authors acknowledge the support of NATO-CP Advanced Fellowship Programme of T\"{U}BITAK (Turkish Scientific Research Council).
V.T. was partially supported by the Program for the leading scientific schools (\# 5747.2006.1), Russian Foundation for Basic Research (\# 05-01-00567) and the Program of Russian Government for the support of National Universities.

\end{document}